\documentclass{jpsj2}

\usepackage{graphicx,amsmath, bm}

\def\d{\mathrm{d}}
\def\e{\mathrm{e}}
\def\i{\mathrm{i}}
\def\p{\varphi}
\def\r{\rho}
\def\P{\Pi}

\def\U{\mathcal{U}}
\def\tr{{\rm tr}}
\def\sol{\mathfrak{D}}

\title{Dark Solitons in $F=1$ Spinor Bose--Einstein Condensate }

\author{Masaru \textsc{Uchiyama}$^{1}$
\thanks{E-mail address: uchiyama@monet.phys.s.u-tokyo.ac.jp}, 
Jun'ichi \textsc{Ieda}$^{2}$\thanks{E-mail address: ieda@imr.tohoku.ac.jp} 
and Miki \textsc{Wadati}$^{1}$}

\inst{$^{1}$Department of Physics, Graduate School of Science,
University of Tokyo, Tokyo 113-0033 \\
$^{2}$Institute for Materials Research, Tohoku University, Sendai, 980-8577}

\abst{We study dark soliton solutions of a multi-component Gross--Pitaevskii
equation for hyperfine spin $F=1$ spinor Bose--Einstein condensate. 
The interactions are supposed to be inter-atomic repulsive and anti-ferromagnetic ones
of equal magnitude. 
The solutions are obtained from those of an integrable $2\times 2$ matrix nonlinear 
Schr\"{o}dinger equation with nonvanishing boundary conditions. 
We investigate the one-soliton and two-soliton solutions in detail. 
One-soliton is classified into two kinds. 
The ferromagnetic state has wavefunctions of domain-wall shape and 
its total spin is nonzero. 
The polar state provides a hole soliton and its total spin is zero.
These two states are selected by choosing the type of the boundary conditions. 
In two-soliton collisions, we observe the spin-mixing or spin-transfer. 
It is found that, as ``magnetic" carriers, 
solitons in the ferromagnetic state are operative for the spin-mixing 
while those in the polar are passive. 
}

\kword{Bose--Einstein condensate, integrable, 
nonlinear Schr\"{o}dinger equation, multi-component, dark soliton, magnetic property}

\begin{document}
\maketitle
\newpage

\section{Introduction} 

Bose--Einstein condensate (BEC) with internal degrees of freedom has been much studied 
theoretically and experimentally. 
Bose--Einstein condensation of ultra-cold bosonic atoms has been realized under 
optical dipole traps \cite{Stenger,Stamper,Miesner}. 
This development stimulates a further research on physical properties of 
the condensate. Moreover, applications in atom optics \cite{Meystre} raise their 
potential, such as atom laser, atom interferometry and coherent atom transport. 
Internal degrees of 
freedom of atoms which are frozen under magnetic trap are liberated and play a 
significant role with wide possibilities in the condensate under optical traps. 

For BEC of dilute gases, the mean-field theory is valid. The time-evolution of the 
condensate is described by the Gross--Pitaevskii (GP) equation. It is known that 
the GP equation for single-component BEC is integrable 
in the one-dimensional (1D) uniform systems and has soliton solutions. 
In experiments, matter-wave dark and bright solitons are produced for single-component 
BECs \cite{Burger,Denschlag,Strecker,Khaykovich}. 

In general, we may suppose a system where each bosonic atom has 
the hyperfine spin of integer $F$. 
The condensate in the spin $F$ 
state is described by the $(2F+1)$-component macroscopic wavefunction. 
Time-evolution of the multi-component condensate is derived from a generalized GP
functional. The obtained multi-component GP equation should explain 
a variety of static and dynamical phenomena in the condensate. 
In particular, the existence of multi-component solitons is expected in the system. 

To analyze such nonlinear evolution equations exactly, we concentrate on the 
one-dimensional $F=1$ spinor condensate. Recently, we have discovered a novel 
reduction to the integrable model \cite{IMW1,IMW2}. There, the atomic interaction is 
attractive and the spin-exchange interaction is ferromagnetic. 
In this paper, we consider the case of the repulsive atomic interaction and 
the anti-ferromagnetic spin-exchange interaction. 
We investigate in detail the one-soliton and two-soliton solutions and 
clarify their magnetic properties, which are absent in single-component solutions. 
We emphasize that the systematic study of 
multi-component soliton equations is rather new and has clarified 
interesting collision properties of solitons \cite{TW1,T}. 

The paper consists of the following. In $\S$~\ref{sec:spinor BEC}, we formulate the 
Gross--Pitaevskii equation for $F=1$ spinor condensate, and present an integrable 
model with repulsive and anti-ferromagnetic interactions. The model is a reduction 
of a $2\times 2$ matrix nonlinear Schr\"{o}dinger equation. 
In $\S$~\ref{sec:soliton solution}, solutions under constant boundary conditions are 
obtained by use of the inverse scattering method. 
Needless to say that the inverse scattering method for constant boundary conditions 
are much more involved than that for vanishing boundary conditions. 
The detail of the inverse scattering method is reported in a separated paper \cite{IUW}. 
One-soliton states and two-soliton 
states are analyzed in detail respectively in $\S$~\ref{sec:1soliton} 
and $\S$~\ref{sec:2soliton}. The last section is devoted to conclusion. 
The generic two-soliton solution is explicitly given in Appendix.

\section{$F=1$ Spinor Bose--Einstein Condensate}
\label{sec:spinor BEC}

Bosonic atoms in the $F=1$ hyperfine state are 
expressed as a three-component quantum field 
operator $\hat{\bm{\Psi}}=(\hat{\Psi}_1, \hat{\Psi}_0, \hat{\Psi}_{-1})$ whose 
equal-time commutation relations are 
\begin{align}
[\hat{\Psi}_\alpha(x,t), \hat{\Psi}_\beta^\dagger(x',t)]=
\delta_{\alpha\beta}\delta(x-x'),
\end{align}
for $\alpha,\beta=1,0,-1$. 
The interaction between atoms is supposed to be of short-range and have a form,  
\begin{align}
\hat{V}(x_1-x_2)=\delta(x_1-x_2)(c_0+c_2\hat{\bm{F}_1}\cdot\hat{\bm{F}_2}),
\end{align}
where $\hat{\bm{F}_i}$ is the spin operator. 
Explicitly, the coupling constants are 
\begin{align}
&c_0=4\pi\hbar^2(a_0+2a_2)/3m,\\
&c_2=4\pi\hbar^2(a_2-a_0)/3m,
\end{align}
where $a_f$ is the $s$-wave scattering length for the channel of total hyperfine spin 
$f$ and $m$ is the mass of the atom. 
$f=2$ corresponds to parallel spin collision and $f=0$ to antiparallel spin collision. 
In this paper, we assume the system is quasi-one dimensional.
The effective 1D couplings are 
\begin{align}
\bar{c}_0=c_0/2a_\perp^2,\qquad \bar{c}_2=c_2/2a_\perp^2,
\end{align}
where $a_\perp$ is the size of the transverse ground state. 
The second quantized Hamiltonian is 
\begin{align}
H=&\int \d x \,\bigg(\frac{\hbar^2}{2m}\partial_x\hat{\Psi}^\dagger_\alpha
\cdot\partial_x\hat{\Psi}_\alpha+
\frac{\bar{c}_0}{2}\hat{\Psi}^\dagger_\alpha
\hat{\Psi}^\dagger_{\alpha^\prime} \hat{\Psi}_{\alpha^\prime}
\hat{\Psi}_\alpha \nonumber \\
&\quad\quad\quad{}+\frac{\bar{c}_2}{2}\hat{\Psi}^\dagger_\alpha
\hat{\Psi}^\dagger_{\alpha^\prime}\bm{\mathsf{f}}_{\alpha \beta} \cdot
\bm{\mathsf{f}}_{\alpha^\prime \beta^\prime}\hat{\Psi}_{\beta^\prime}
\hat{\Psi}_{\beta}\bigg).
\end{align}
As usual, the repeated indices mean the summation. 
Explicit form is the following, 
\begin{align}
H=&\int \d x\, \bigg\{\frac{\hbar^2}{2m}\partial_x\hat{\Psi}^\dagger_\alpha
\cdot\partial_x\hat{\Psi}_\alpha+\frac{\bar{c}_0+\bar{c}_2}{2}
\Big[\hat{\Psi}^\dagger_1 \hat{\Psi}^\dagger_1
\hat{\Psi}_1 \hat{\Psi}_1 \nonumber\\
&{}+\hat{\Psi}^\dagger_{-1}
\hat{\Psi}^\dagger_{-1} \hat{\Psi}_{-1} \hat{\Psi}_{-1}
+2\hat{\Psi}^\dagger_0 \hat{\Psi}_0(\hat{\Psi}^\dagger_{1}
\hat{\Psi}_1+\hat{\Psi}^\dagger_{-1}\hat{\Psi}_{-1})\Big]
\nonumber\\
&{}+(\bar{c}_0-\bar{c}_2) \hat{\Psi}^\dagger_1 \hat{\Psi}^\dagger_{-1}
\hat{\Psi}_{-1} \hat{\Psi}_1
+\frac{\bar{c}_0}{2}\hat{\Psi}^\dagger_0 \hat{\Psi}^\dagger_0
\hat{\Psi}_0 \hat{\Psi}_0
\nonumber\\
&{}+\bar{c}_2 \left[\hat{\Psi}^\dagger_1 \hat{\Psi}^\dagger_{-1}
\hat{\Psi}_0 \hat{\Psi}_0+\hat{\Psi}^\dagger_0 \hat{\Psi}^\dagger_0
\hat{\Psi}_{-1} \hat{\Psi}_1\right]\bigg\},
\end{align}
where we have used 
\begin{align}
&\mathsf{f}^x=\frac{1}{\sqrt{2}}\left(\hspace*{-1mm}\begin{array}{ccc}
0\hspace*{-1mm}&\hspace*{-1mm}1\hspace*{-1mm}&\hspace*{-1mm}0 \\
1\hspace*{-1mm}&\hspace*{-1mm}0\hspace*{-1mm}&\hspace*{-1mm}1 \\
0\hspace*{-1mm}&\hspace*{-1mm}1\hspace*{-1mm}&\hspace*{-1mm}0
\end{array}\hspace*{-1mm}\right),\ \ 
\mathsf{f}^y=\frac{\i}{\sqrt{2}}\left(\hspace*{-1mm}\begin{array}{ccc}
0\hspace*{-1mm}&\hspace*{-1mm}-1\hspace*{-1mm}&\hspace*{-1mm} 0 \\
1\hspace*{-1mm}&\hspace*{-1mm} 0\hspace*{-1mm}&\hspace*{-1mm}-1 \\
0\hspace*{-1mm}&\hspace*{-1mm} 1\hspace*{-1mm}&\hspace*{-1mm} 0
\end{array}\hspace*{-1mm}\right),
\nonumber\\
&\mathsf{f}^z=\left(\hspace*{-1mm}\begin{array}{ccc}
1\hspace*{-1mm}&\hspace*{-1mm}0\hspace*{-1mm}&\hspace*{-1mm} 0 \\
0\hspace*{-1mm}&\hspace*{-1mm}0\hspace*{-1mm}&\hspace*{-1mm} 0 \\
0\hspace*{-1mm}&\hspace*{-1mm}0\hspace*{-1mm}&\hspace*{-1mm}-1
\end{array}\hspace*{-1mm}\right).
\label{matrixf}
\end{align}
In the mean-field theory of BEC, the dynamical variable $\bm{\Phi}(x,t)$ is 
the vacuum expectation value of the quantum field, 
\begin{align}
{\bm\Phi}(x,t)&= \langle \hat{\bm \Psi}(x,t) \rangle\nonumber\\
&=\left(\Phi_1(x,t),\Phi_0(x,t),\Phi_{-1}(x,t)\right)^T,
\end{align}
which is normalized to the total number of atoms $N_T$, 
\begin{align}
\int \d x \,{\bf \Phi}^{\dagger}(x,t)  \cdot {\bf \Phi}(x,t)
=N_T.
\end{align}
The time-evolution of the spinor condensate wavefunction ${\bm \Phi}(x,t)$
can be derived from the variational principle:
\begin{align}
\label{variation}
\i\hbar\partial_t \Phi_\alpha(x,t)=\frac{\delta E_{\rm GP}}
{\delta \Phi^*_\alpha(x,t)},
\end{align}
where the Gross--Pitaevskii energy functional is given by
\begin{align}
\label{MFenergy}
E_{\rm GP}&=\int \d x\, \bigg\{\frac{\hbar^2}{2m}\partial_x\Phi^*_\alpha
\cdot\partial_x\Phi_\alpha
+\frac{\bar{c}_0+\bar{c}_2}{2}\Big[|\Phi_1|^4+|\Phi_{-1}|^4\nonumber\\
&{}+2|\Phi_0|^2(|\Phi_1|^2+|\Phi_{-1}|^2)\Big]
+(\bar{c}_0-\bar{c}_2)|\Phi_1|^2|\Phi_{-1}|^2
\nonumber\\
&{}+\frac{\bar{c}_0}{2}|\Phi_0|^4+\bar{c}_2(\Phi^*_1\Phi^*_{-1}\Phi^2_0
+{\Phi^*_0}^2\Phi_1\Phi_{-1})\bigg\}.
\end{align}
Substituting eq.(\ref{MFenergy}) into eq.(\ref{variation}),
we get a set of nonlinear evolution equations for the spinor condensate
wavefunctions:
\begin{align}
\i\hbar\partial_t\Phi_1=&
 -\frac{\hbar^2}{2m}\partial^2_x\Phi_1+
(\bar{c}_0+\bar{c}_2)(|\Phi_1|^2+|\Phi_{0}|^2)\Phi_1\nonumber\\
&{}+(\bar{c}_0-\bar{c}_2)|\Phi_{-1}|^2\Phi_1
+\bar{c}_2\Phi^*_{-1}\Phi^2_0,\nonumber\\
\i\hbar\partial_t\Phi_0=&
-\frac{\hbar^2}{2m}\partial^2_x\Phi_0+(\bar{c}_0+\bar{c}_2)
(|\Phi_1|^2+|\Phi_{-1}|^2)\Phi_0\nonumber\\
&{}+\bar{c}_0|\Phi_0|^2\Phi_0
+2\bar{c}_2\Phi^*_0\Phi_1\Phi_{-1},\nonumber\\
\label{TDeq-1}
\i\hbar\partial_t\Phi_{-1}=&
-\frac{\hbar^2}{2m}\partial^2_x\Phi_{-1}
+(\bar{c}_0+\bar{c}_2)(|\Phi_{-1}|^2+|\Phi_{0}|^2)\Phi_{-1}\nonumber\\
&{}+(\bar{c}_0-\bar{c}_2)|\Phi_{1}|^2\Phi_{-1}
+\bar{c}_2\Phi^*_{1}\Phi^2_0.
\end{align}
We call eqs.(\ref{TDeq-1}) multi-component Gross--Pitaevskii (GP) equation 
for $F=1$ spinor Bose--Einstein condensate. 

Now, we consider a system with couplings $\bar{c}_0=\bar{c}_2=c>0$. 
This corresponds to the situation 
with repulsive inter-atomic interaction and anti-ferromagnetic 
spin-exchange interaction whose magnitudes are equal. 
In the dimensionless form:
${\bm \Phi}\to (\phi_1, \sqrt{2}\phi_0,\phi_{-1})^T$,
where time and length are measured respectively
in units of $\bar{t}=\hbar a_\perp/c$ and $\bar{x}=\hbar\sqrt{a_\perp/2mc}$,
we rewrite eqs.(\ref{TDeq-1}) as follows,
\begin{align}
&\i\partial_t \phi_1= -\partial^2_{x}\phi_1
+2(|\phi_1|^2+2|\phi_0|^2)\phi_1+2\phi^*_{-1}\phi^2_0,\nonumber\\
&\i\partial_t \phi_0= -\partial^2_{x}\phi_0
+2(|\phi_{-1}|^2+|\phi_0|^2+|\phi_1|^2)\phi_0+2\phi^*_0\phi_1\phi_{-1},
\nonumber\\
&\i\partial_t \phi_{-1}= -\partial^2_{x}\phi_{-1}
+2(|\phi_{-1}|^2+2|\phi_0|^2)\phi_{-1}+2\phi^*_{1}\phi^2_0.
\end{align}
These coupled equations are equivalent to a $2 \times 2$ matrix version of
nonlinear Schr\"{o}dinger equation (NLSE) with the self-defocusing nonlinearity:
\begin{align}
\label{NLS}
\i\partial_t Q+ \partial^2_x Q-2QQ^\dagger Q=O,
\end{align}
with an identification,
\begin{align}
\label{reduction}
Q=\left(\hspace*{-1mm}\begin{array}{cc}
\phi_1 \hspace*{-1mm}&\hspace*{-1mm} \phi_0 \\
\phi_0 \hspace*{-1mm}&\hspace*{-1mm} \phi_{-1}
\end{array}\hspace*{-1mm}
\right).
\end{align}
Two remarks are in order. First, eq.(\ref{NLS}) is a completely integrable system. 
Second, for unitary matrices $\mathcal{U}$ and $\mathcal{V}$, 
if $Q$ is a solution of eq.(\ref{NLS}), then 
$\mathcal{U}Q\mathcal{V}$ is also a solution.

\section{Solitons of $2\times 2$ Matrix NLSE with Nonvanishing Boundary Conditions}
\label{sec:soliton solution}

In a separate paper~\cite{IUW}, we investigated the inverse scattering method (ISM) 
for the self-defocusing matrix NLSE with nonvanishing boundary conditions. 
It is shown that the standard form of the $N$-soliton solution is expressed as 
\begin{align}
\label{Nsoliton}
Q(x,t)=\lambda_0 \e^{\i\phi(x,t)}
\left[ I+2\i (\overbrace{I\cdots I}^N) S^{-1} \left(
\begin{array}{c}
\P_1 \e^{\chi_1}\\
\vdots\\
\P_N \e^{\chi_N}
\end{array}
\right) \right].
\end{align}
Here $I$ is the $2\times 2$ unit matrix, and
$\P_i$ ($i=1,\cdots,N$) are $2\times 2$ Hermitian matrices, 
called the polarization matrices. 
$S$ is a $2N\times 2N$ matrix defined by 
\begin{align}
S_{ij}=\e^{-\i\p_{i}}\delta_{ij}I+\frac{\e^{-\i\p_i}+\e^{-\i\p_j}}{\e^{-\r_i}+\e^{-\r_j}}
\P_i \e^{\chi_i},\\
1\le i,j\le N, \nonumber
\end{align}
where $\p_j\in(0,\pi)$ and $\e^{-\r_j}=\sin\p_j$ for $j=1,\cdots,N$. 
The phase of the carrier wave is given by 
\begin{align}
\phi(x,t)=kx-(k^2+2\lambda_0^2)t+\delta,
\end{align}
and the coordinate function for the $j$-th soliton is given by 
\begin{align}
\chi_j(x,t)=-2\lambda_0\sin\p_j(x-2(\lambda_0\cos\p_j+k)t).
\end{align}
Initial position of the $j$-th soliton is adjusted by scaling $\P_j$. 
The analysis of the ISM shows that we should take $\det \P_j=0$ for all $j$. 
In that case, however, multi-soliton with the same spectral parameters turns to be 
one-soliton and 
is represented by a single polarization matrix with nonzero determinant (see Appendix). 
The expression (\ref{Nsoliton}) holds true even if this type of degenerate 
multi-solitons are included. 
Therefore, we relax the condition so that $\det\P_j$ can have nonzero values. 
To repeat, $Q(x,t)$ with arbitrary real $\det\P_j$ in general is 
the $N$-soliton solution of eq.(\ref{NLS}). 

The solution (\ref{Nsoliton}) has a form very close to the bright soliton solutions for 
the self-focusing matrix NLSE. However, there exists a significant difference: 
since the boundary conditions are nonvanishing 
and the nonlinearity is self-defocusing, the solitons are dark solitons, in general. 
The boundary condition in the limit $x\to\infty$ is fixed to be 
\begin{align}
Q(x,t)\e^{-\i\phi(x,t)}\to \lambda_0 I,\qquad x\to\infty.
\end{align}
The limit $x\to -\infty$ depends on whether $\det\P_j=0$ or not, which will be
discussed later for the one-soliton case and the two-soliton case. 
Other types of boundary conditions can be also realized from this standard solution 
by unitary transformations. 

For our setting of $F=1$ spinor BEC, the explicit form of the $N$-soliton 
solution is obtained 
through the identification (\ref{reduction}) in eq.(\ref{Nsoliton}). 
Noting that $Q$ in eq.(\ref{reduction}) is a symmetric matrix, we naturally take $\P_j$ 
to be real symmetric. 
An SU(2) transformation of a spinor wavefunction, 
$\bm{\Phi}'=U\bm{\Phi}$ with $U=\exp[\i\bm{\theta}\cdot\bm{\mathsf{f}}]$, 
is equivalent to the following unitary transformation of $Q$: 
\begin{align}
\label{rotation}
Q'=\U Q\U^T, \qquad \U=\exp[\i\bm{\theta}\cdot\bm{\sigma}/2],
\end{align}
where $\bm{\mathsf{f}}=(\mathsf{f}^x, \mathsf{f}^y, \mathsf{f}^z)$ are 
defined in eq.(\ref{matrixf}) and 
$\bm{\sigma}=(\sigma^x, \sigma^y, \sigma^z)$ are the Pauli matrices. 
By this transformation, the boundary condition in the limit $x\to\infty$ 
can take any values except $\phi_1=0=\phi_{-1}$. 
For the exceptional case $\phi_1=0=\phi_{-1}$, we may identify $Q$ instead by 
\begin{align}
Q=\left(\hspace*{-1mm}\begin{array}{cc}
\phi_0 \hspace*{-1mm}&\hspace*{-1mm} \phi_1 \\
\phi_{-1} \hspace*{-1mm}&\hspace*{-1mm} \phi_0
\end{array}\hspace*{-1mm}
\right).
\end{align}
Solitons in this case are just reduced to the one-component ones; 
$\phi_1=0=\phi_{-1}$ everywhere. That is, we realize 
one-component dark soliton solutions from the three-component solitons. 
Seen as a solution of the matrix NLSE, this corresponds to the degenerate one. 
For example, a twin-peak of the hole number density collapses to a single-peak. 
We shall discuss this situation in \S\ref{sec:polar}. 

The matrix NLSE has infinitely many conserved quantities due to its complete 
integrability~\cite{TW1}. 
First some conserved quantities are related to physical quantities as listed below: 
\begin{align}
&\mbox{\textbf{Total hole number:} }\quad \bar{N_T}=\int \d x\ \bar{n}(x,t),\\
&\qquad\bar{n}(x,t)=\tr(Q_\pm^\dagger Q_\pm)-\tr(Q^\dagger Q).\\
&\mbox{\textbf{Total spin:} }\quad \bm{F}_T=\int \d x\ \bm{f}(x,t),\\
&\qquad\bm{f}(x,t)=\tr(Q^\dagger \bm\sigma Q).\\
&\mbox{\textbf{Total hole momentum:} }\quad \bar{P_T}=\int \d x\ \bar{p}(x,t),\\
&\qquad\bar{p}(x,t)=-\i\hbar[\tr(Q_\pm^\dagger Q_{\pm,x})-\tr(Q^\dagger Q_x)].\\
&\mbox{\textbf{Total hole energy:} }\quad \bar{E_T}=\int \d x\ \bar{e}(x,t),\\
&\qquad\bar{e}(x,t)=
	c[\tr(Q_{\pm,x}^\dagger Q_{\pm,x}+Q_\pm^\dagger Q_\pm Q_\pm^\dagger Q_\pm)\nonumber\\
&\hspace{3cm}-\tr(Q_x^\dagger Q_x+Q^\dagger QQ^\dagger Q)].
\end{align}
Since there remain carrier waves at infinity, 
meaningful finite quantities are calculated 
by subtracting the background $Q_\pm=\lim_{x\to\pm\infty}Q(x)$. 
The bar denotes such a hole contribution calculated from the background. 
Physical quantities without the bar express particle properties. 
Among local quantities, 
the spin density $\bm{f}=(f_x, f_y, f_z)$ is 
covariant under SU(2) transformation (\ref{rotation}) and rotated as 
\begin{align}
\label{spinrotation}
\bm{f}'=\mathcal{R}\bm{f}=\tr(Q^\dagger\U^\dagger\bm{\sigma}\U Q).
\end{align}
If we choose $\U=\exp[\i\gamma\sigma^z/2]\exp[\i\beta\sigma^x/2]
\exp[\i\alpha\sigma^z/2]$ ($\alpha, \beta, \gamma$: Euler angles), 
the spin-rotation corresponds to 
\begin{align}
\mathcal{R}=\mathcal{R}(\alpha,\beta,\gamma)\equiv
\mathcal{R}_z(-\gamma)\mathcal{R}_x(-\beta)\mathcal{R}_z(-\alpha),
\end{align}
where $\mathcal{R}_i(\xi)$ ($i=x,y,z$) is the matrix representing the rotation 
of an angle $\xi$ around the $i$-th axis. 
The other densities such as $\bar{n}(x,t)$, $\bar{p}(x,t)$ and $\bar{e}(x,t)$ 
are invariant under SU(2) transformation. 
The SU(2) symmetry of the system is attributed to the energy degeneracy 
for this spin rotation. 
The appearance of the spin density is the most remarkable feature for 
multi-component solitons.

\section{One-Soliton States}
\label{sec:1soliton}

In this section, we classify the one-soliton solution for our $F=1$ spinor BEC. 
Setting $N=1$ in eq.(\ref{Nsoliton}), the standard form is expressed as 
\begin{align}
\label{eq:1soliton}
&Q(x,t)=\lambda_0 \e^{\i\phi(x,t)}\sol(\chi;\p,\P)\nonumber\\
&\equiv\lambda_0 \e^{\i\phi(x,t)}
\frac{(\e^{2(\chi+\r+\i\p)}\det\P+\e^{\chi+\r}\tr\P+1)I+2\i\e^{\chi+\i\p}\P}
{\e^{2(\chi+\r)}\det\P+\e^{\chi+\r}\tr\P+1},
\end{align}
where we put $\e^{-\r}=\sin\p$. 
Again, the phase of the carrier wave is given by 
\begin{align}
\phi(x,t)=kx-(k^2+2\lambda_0^2)t+\delta,
\end{align}
and the coordinate of the envelop-soliton is given by 
\begin{align}
\chi(x,t)=-2\lambda_0\sin\p(x-2(\lambda_0\cos\p+k)t).
\end{align}
In eq.(\ref{eq:1soliton}), 
$\P$ is a real symmetric $2\times 2$ matrix. In order to have a non-singular solution, 
it is necessary that $\det\P\ge 0$ and $\tr\P>0$. 

There are 3 parameters and a matrix for a one-soliton: 
\begin{center}
\begin{description}
\item \qquad$k:$ wave number of carrier wave,
\item \qquad$\lambda_0:$ amplitude of carrier wave,
\item \qquad$\p:$ the group velocity $2(\lambda_0\cos\p+k)$ and\\
\qquad the inverse of the size of soliton $2\lambda_0\sin\p$,
\item \qquad$\P:$ contributions of the spin components.
\end{description}
\end{center}
As is often the case for solitons, the amplitude determines the speed, and vice versa. 
The polarization matrix $\P$ is a new ingredient for multi-component solitons and 
actually there is no counterpart in one-component solitons. 
The densities are explicitly calculated as follows, 
\begin{align}
&\bar{n}(x,t)=\frac{4\lambda_0^2\e^{\chi-\r}}{(\det S)^2} 
(\e^{2\chi+2\r} \det\P \ \tr\P +4 \e^{\chi+\r} \det\P +\tr\P ),\\
&\bm{f}(x,t)=\frac{4\lambda_0^2 \e^{\chi-\r}\tr\P \bm{\sigma}}{(\det S)^2} 
(\e^{2\chi+2\r} \det\P -1),\\
&\bar{p}(x,t)=\hbar(k+\lambda_0\cos\p)\bar{n},
\end{align}
where $\det S=\e^{2\chi+2\r}\det\P +\e^{\chi+\r}\tr \P+1$. 

We have two kinds of solutions corresponding to $\det\P=0$ and $\det\P>0$, respectively. 
For $\det\P=0$, the total spin is nonzero, 
while for $\det\P>0$, the total spin is zero. 
Other than the total spin, the boundary conditions are also different for these cases. 
Therefore, we can distinguish one-solitons by checking the boundary values. 
We detail these two solutions in the next subsections.

\subsection{Ferromagnetic state}
Consider the case where $\det\P=0$. 
We can write the symmetric matrix $\P$ as 
\begin{align}
\P =\Lambda \left(\hspace*{-1mm}
\begin{array}{cc}
\cos^2\frac{\theta}{2} \hspace*{-1mm}& 
\hspace*{-1mm}\sin\frac{\theta}{2}\cos\frac{\theta}{2}\\
\sin\frac{\theta}{2}\cos\frac{\theta}{2}\hspace*{-1mm} &
\hspace*{-1mm}\sin^2\frac{\theta}{2}
\end{array}
\hspace*{-1mm}\right),
\end{align}
with a non-zero constant $\Lambda$. 
The boundary conditions of the standard form (\ref{eq:1soliton}) are 
\begin{align}
Q\e^{-\i\phi}\to \lambda_0 I
&,\qquad x\to\infty,\\
Q\e^{-\i\phi}\to 
\lambda_0\ \U\U^T
&,\qquad x\to-\infty,
\end{align}
where 
\begin{align}
\U=\e^{\i\p/2} \exp[-\i\theta\sigma^y/2]\exp[\i\p\sigma^z/2].
\label{boundaryU}
\end{align}
Comparing the left and right boundary values, we see that, 
not only the global phase but also the amplitude of each component are different.
That is, the SU(2) rotated boundary conditions. 
It is remarkable that the profiles of the wavefunctions are 
in the shape of domain-walls. 
In contrast, the particle number density profiles have a dark soliton shape. 
See the upper panels of Fig.\ref{fig:1soliton}. 
In particular, the spin is nonzero in total. Therefore, we call the soliton being in 
the \textit{ferromagnetic state}. 

Conserved quantities are integrated as follows. 
\begin{align}
&\bar{N_T}=2\lambda_0\sin\p ,\\
&\bm{F}_T=-\bar{N_T}\frac{\tr\P \bm{\sigma}}{\tr\P}
=-\bar{N_T}(\sin\theta,0,\cos\theta)^T, \label{eq:ferrototalspin}\\
&\bar{P_T}=\bar{N_T}\hbar(k+\lambda_0\cos\p),
\end{align}
\begin{align}
\bar{E_T}&=\bar{T}+\bar{V}\nonumber\\
&=\bar{N_T}c\left((k+\lambda_0\cos\p)^2+\lambda_0^2
-\frac{1}{3}\lambda_0^2\sin^2\!\p\right),\\
\bar{T}&=\bar{N_T}c\left((k+\lambda_0\cos\p)^2-\lambda_0^2+
\frac{1}{3}\lambda_0^2\sin^2\!\p\right),\\
\bar{V}&=\bar{N_T}c\left(2\lambda_0^2-\frac{2}{3}\lambda_0^2\sin^2\!\p\right).
\end{align}
Changing the boundary conditions by global SU(2) transformation, 
we can vary the direction of the spin density as shown in eq.(\ref{spinrotation}).

\subsection{Polar state}
\label{sec:polar}
\begin{figure}[htbp]
\unitlength=1mm
\begin{picture}(180,75)
\put(18,-3){(a)}
\put(73,-3){(b)}
\put(130,-3){(c)}
\hspace*{-1cm}
\begin{minipage}[b]{55mm}
\includegraphics[scale=.40]{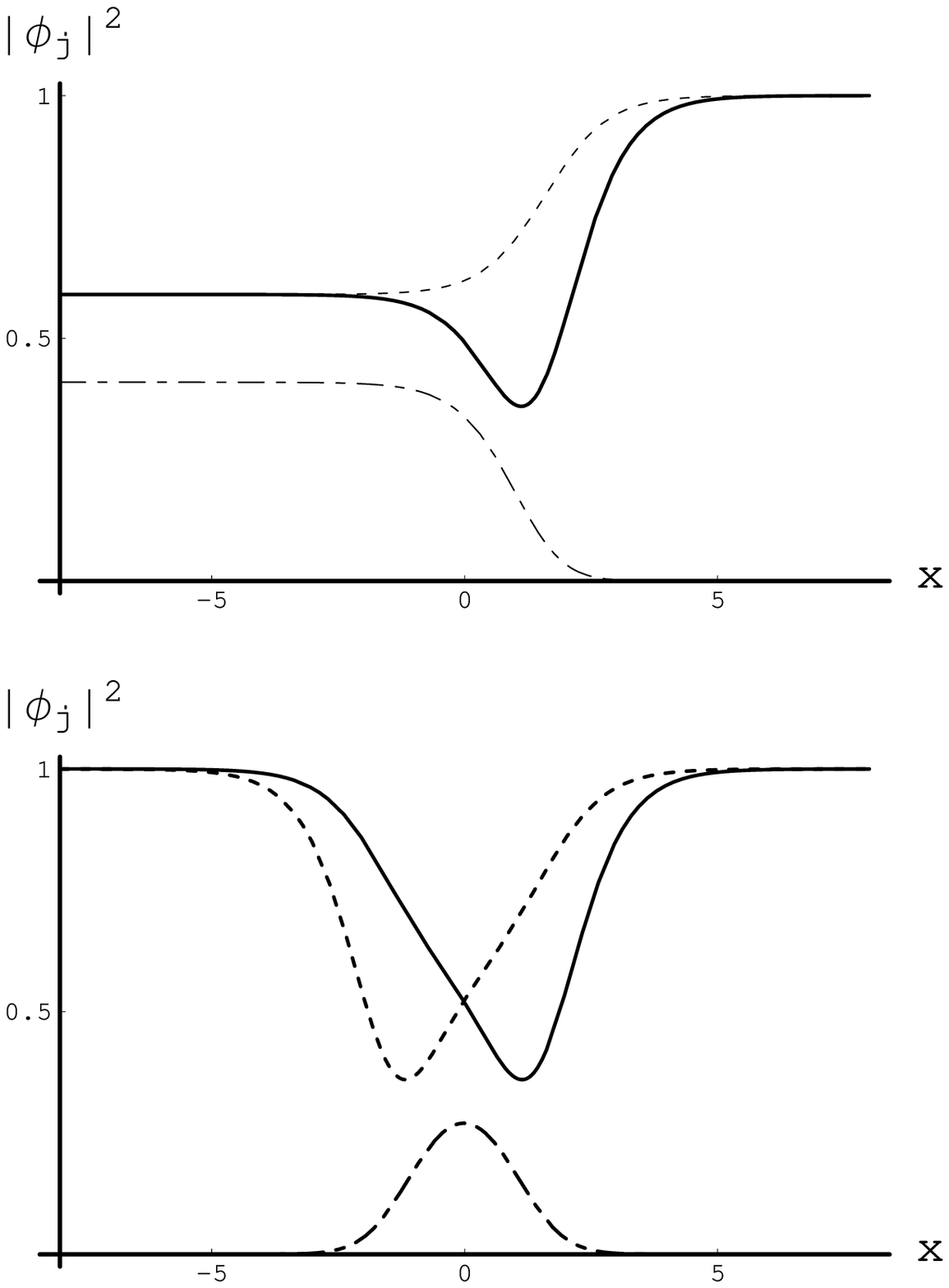}
\end{minipage}
\begin{minipage}[b]{55mm}
\includegraphics[scale=.40]{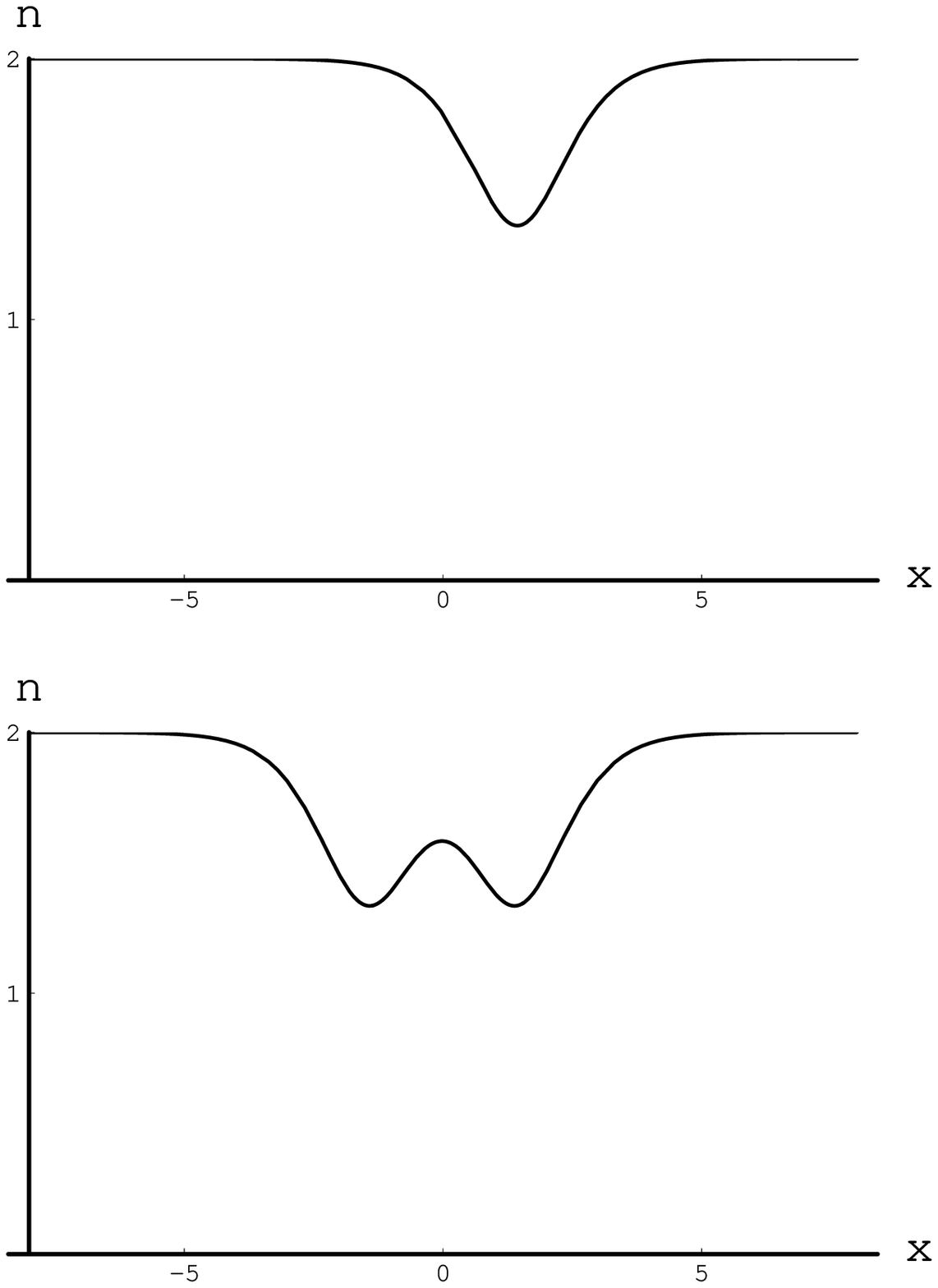}
\end{minipage}
\begin{minipage}[b]{55mm}
\includegraphics[scale=.40]{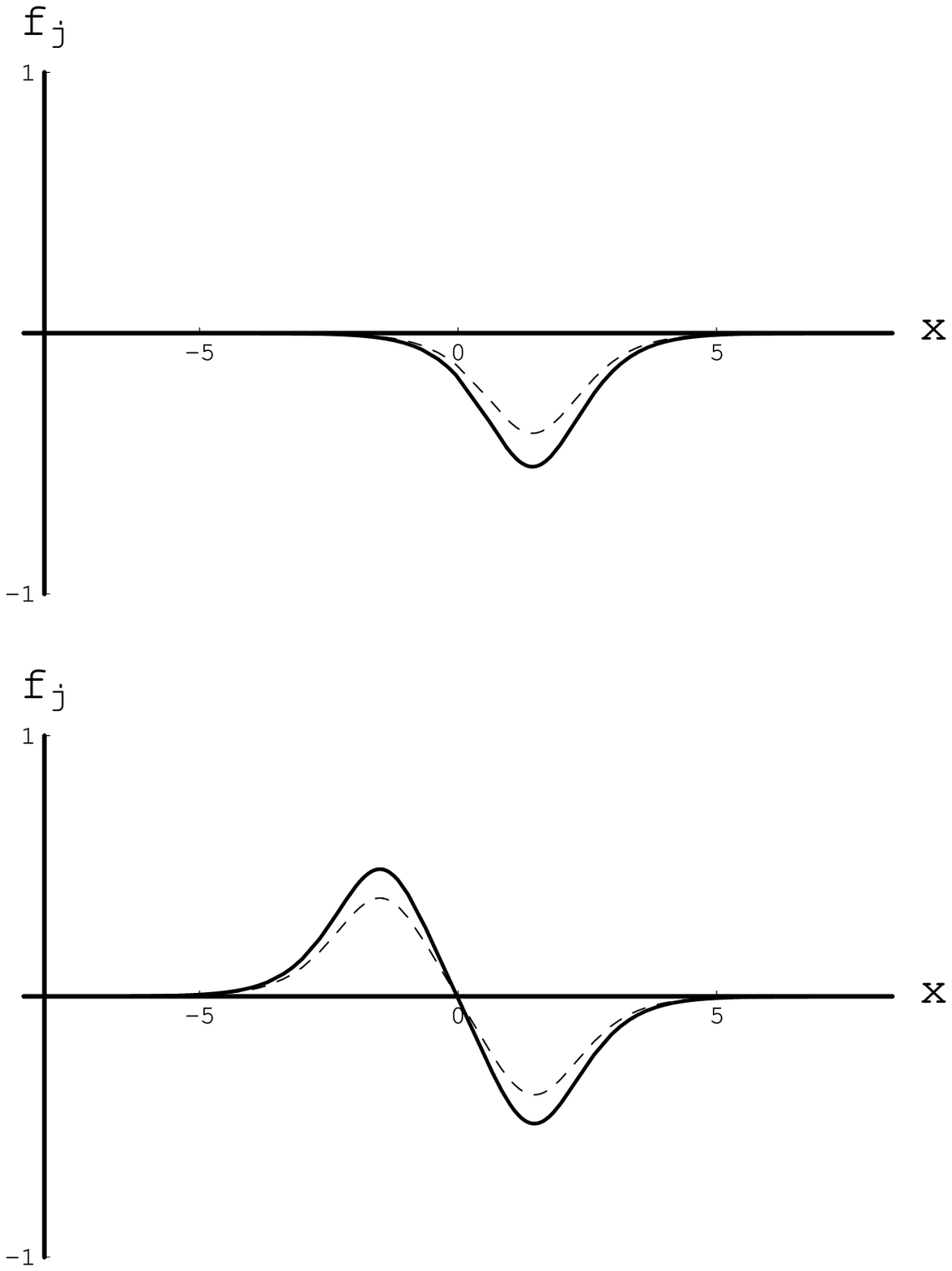}
\end{minipage}
\end{picture}
\caption{Snapshots of $1$-soliton density profiles. 
The upper panels show plots for the ferromagnetic state with 
$\lambda_0=1,\ k=0,\ \p=0.927,\ 
\P=\left(\hspace*{-2mm}
\begin{array}{cc}
8\hspace*{-1mm}\vspace{-1mm}&\hspace*{-1mm}4\\ 
4\hspace*{-1mm}&\hspace*{-1mm}2
\end{array}
\hspace*{-2mm}\right)$. 
The lower panels show plots for the polar state with 
$\lambda_0=1,\ k=0,\ \p=0.927,\ 
\P=\left(\hspace*{-3mm}
\begin{array}{cc}
8\hspace*{-2mm}\vspace{-1mm}&\hspace*{-2mm}3.88\\ 
3.88\hspace*{-2mm}&\hspace*{-2mm}2
\end{array}
\hspace*{-3mm}\right)$. 
(a) density of each component, $|\phi_1|^2$ (solid line), 
$|\phi_0|^2$ (chain line) and $|\phi_{-1}|^2$ (dotted line), (b) particle number density $n$, 
(c) spin densities, $f_x$ (solid line) and $f_z$ (dashed line). 
$f_y$ is zero everywhere for both cases. 
}
\label{fig:1soliton}
\end{figure}

Consider the case where $\det\P>0$. 
In fact, the one-soliton with $\det\P>0$ is obtained as a multi-ferromagnetic-soliton 
solution where $\det\P_j$=0 and $\p_j$ are set the same value for all $j$. 
In spite of the multi-soliton content, it behaves as a one-soliton. 
The boundary conditions of the standard form (\ref{eq:1soliton}) are 
\begin{align}
Q\e^{-\i\phi}\to \lambda_0 I&,\qquad x\to\infty,\\
Q\e^{-\i\phi}\to \lambda_0 \e^{2\i\p}I&,\qquad x\to-\infty.
\end{align}
Unlike the ferromagnetic state, boundary values differ only by phase, 
i.e., U(1) transformation. 
One example of the density profiles is shown in lower panels of Fig.\ref{fig:1soliton}. 
Observing the shape of the profiles, $\phi_{\pm1}$ are dark solitons, i.e. 
hole-like envelopes, and $\phi_0$ is a bright soliton, i.e. a peak. 
The valleys of $\phi_{\pm1}$ are of the same depth but at different positions and 
there is a peak of $\phi_0$ at the midpoint of them.  
The profile of the particle number density has a twin-valley. 
The profile of the spin density is dipole-like shape, i.e. an odd function of the 
coordinate, so that the total spin amounts to zero. 
For this reason, we call the soliton being in the \textit{polar state}. 
By integration, conserved quantities are obtained as follows. 
\begin{align}
&\bar{N_T}=4\lambda_0\sin\p ,\\
&\bm{F}_T=(0,0,0)^T,\\
&\bar{P_T}=\bar{N_T}\hbar(k+\lambda_0\cos\p),
\end{align}
\begin{align}
\bar{E_T}&=\bar{T}+\bar{V}\nonumber\\
&=\bar{N_T}c\left((k+\lambda_0\cos\p)^2+\lambda_0^2
-\frac{1}{3}\lambda_0^2\sin^2\!\p\right),\\
\bar{T}&=\bar{N_T}c\left((k+\lambda_0\cos\p)^2-\lambda_0^2+
\frac{1}{3}\lambda_0^2\sin^2\!\p\right),\\
\bar{V}&=\bar{N_T}c\left(2\lambda_0^2-\frac{2}{3}\lambda_0^2\sin^2\!\p\right).
\end{align}

The twin-valley shape merges when $\det\P=(\tr\P)^2/4$, or equivalently 
$\P\propto I$. 
In this case, the diagonal elements of $Q$ have the same values and the off-diagonal 
elements are zero everywhere. 
In other words, the soliton is reduced to the one-component one. 
In fact, the physical densities are all the same as those of 
the one-component NLSE soliton: 
\begin{align}
&\bar{n}(x,t)=2\lambda_0^2\sin^2\!\p\ \mathrm{sech}^2\left((\chi+\r+\xi)/2\right),\\
&\bm{f}(x,t)=0,\\
&\bar{p}(x,t)=\hbar(k+\lambda_0\cos\p)\bar{n},
\end{align}
where we put $\e^\xi=\tr\P/2$. 
Of course, the spin density is zero everywhere. 

One notices that if the left-half part of the polar soliton in the lower figures of 
Fig.\ref{fig:1soliton} is pulled away to $x=-\infty$, then the remainder is the 
ferromagnetic soliton in the upper figures of Fig.\ref{fig:1soliton}. 
This behavior is the same as 
that observed for the bright polar and ferromagnetic solitons 
in $F=1$ spinor BEC with attractive and ferromagnetic interactions \cite{IMW2}.
In fact, for fixed amplitude and group velocity, the total hole number, 
the total hole momentum and the total hole energy are just twice as 
those of the ferromagnetic soliton. 
It can be said that the polar state is a special case of two ferromagnetic states.

\section{Two-Soliton Collision}
\label{sec:2soliton}
In this section, we analyze the two-soliton solution in detail. 
The standard form of two-soliton solution is given by eq.(\ref{Nsoliton}) with $N=2$. 
The explicit form is very complicated (see Appendix). 

The well-known and most remarkable property of solitons is 
the shape-preserving against mutual collisions. 
However, this is not always true for multi-component solitons in a naive sense. 
We look into the following three cases of two-soliton collisions: 
(i) polar--polar, (ii) ferro--polar, (iii) ferro--ferro. 
We concentrate on the case where soliton $1$ moves rightward faster than soliton $2$; 
soliton 1 will eventually pass over soliton 2. 
Each soliton moves with a constant speed except in the collision region. 
When another soliton is far apart, a soliton almost 
has the form of the one-soliton solution. 
To figure out the asymptotic form of soliton $1$ in $t=\pm\infty$, 
fix the coordinate $\chi_1$ of the envelope of soliton $1$ and then take the limit 
$t\to\pm\infty \Leftrightarrow \e^{\chi_2}\to 0$ or $\infty$. 
Similarly, the asymptotic form of soliton 2 is obtained.

\subsection{Polar state vs. Polar state}
Let us examine the collision of two solitons both in the polar state. 
We set $\det\P_j>0$ for $j=1, 2$. 
Two solitons in each component collide each other in the same manner as the 
one-component two-soliton collision. 
We see that the only change due to the collision is a phase shift. 
Calculating the asymptotic forms from eq.(\ref{2soliton}), 
we have for $t\to-\infty$,
\begin{align}
Q&\simeq Q_1^{\rm in}+Q_2^{\rm in}, \nonumber\\
Q_1^{\rm in}&=\lambda_0\e^{\i\phi+2\i\p_2}\sol\left(\chi_1+s; \p_1, \P_1\right),\\
Q_2^{\rm in}&=\lambda_0\e^{\i\phi}\sol\left(\chi_2; \p_2, \P_2\right),
\end{align}
and for $t\to\infty$,
\begin{align}
Q&\simeq Q_1^{\rm fin}+Q_2^{\rm fin}, \nonumber\\
Q_1^{\rm fin}&=\lambda_0\e^{\i\phi}\sol\left(\chi_1; \p_1, \P_1\right),\\
Q_2^{\rm fin}&=\lambda_0\e^{\i\phi+2\i\p_1}\sol\left(\chi_2+s; \p_2, \P_2\right).
\end{align}
The superscripts in and fin mean initial and final, respectively. 
Recall that $\sol\left(\chi; \p, \P\right)$ is defined in eq.(\ref{eq:1soliton}). 
The position shift is determined by $\e^s=\Xi$ where 
\begin{align}
\Xi=\frac{\sin^2\left(\frac{1}{2}(\p_2-\p_1)\right)}
{\sin^2\left(\frac{1}{2}(\p_2+\p_1)\right)}.
\label{shift}
\end{align}
The boundary conditions of the asymptotic forms are consistent for 
the two solitons. The results show the following: 
the soliton at the left hand is only changed 
by a phase shift and a position shift from 
the standard one-soliton solution in the presence of the soliton at the right hand 
which is in the standard form. 
The conserved quantities are calculated to be the sum of those of two asymptotic 
one-solitons.

\subsection{Ferromagnetic state vs. Polar state}
Let us see the collision between soliton 1 in the ferromagnetic state and 
soliton 2 in the polar state. 
We set $\det\P_1=0$ and $\det\P_2>0$. 
We write the polarization matrix $\P_1$ for soliton 1 as 
\begin{align}
\P _1&=\Lambda_1 \left(\hspace*{-1mm}
\begin{array}{cc}
\cos^2\frac{\theta_1}{2}\hspace*{-1mm} &\hspace*{-1mm} 
\sin\frac{\theta_1}{2}\cos\frac{\theta_1}{2}\\
\sin\frac{\theta_1}{2}\cos\frac{\theta_1}{2}\hspace*{-1mm} &\hspace*{-1mm} 
\sin^2\frac{\theta_1}{2}
\end{array}
\hspace*{-1mm}\right),
\end{align}
and accordingly we write (cf. eq.(\ref{boundaryU}))
\begin{align}
\U_1=\e^{\i\p_1} \exp\left[ -\i\theta_1 \sigma^y/2\right] 
\exp\left[ \i\p_1\sigma^z/2\right].
\end{align}
Then, we calculate the asymptotic forms. The results are for $t\to-\infty$, 
\begin{align}
Q&\simeq Q_1^{\rm in}+Q_2^{\rm in}, \nonumber\\
Q_1^{\rm in}&=\lambda_0\e^{\i\phi+2\i\p_2}\sol\left(\chi_1+s; \p_1, \P_1\right),\\
Q_2^{\rm in}&=\lambda_0\e^{\i\phi}\sol\left(\chi_2; \p_2, \P_2\right),
\end{align}
and for $t\to\infty$, 
\begin{align}
Q&\simeq Q_1^{\rm fin}+Q_2^{\rm fin}, \nonumber\\
Q_1^{\rm fin}&=\lambda_0\e^{\i\phi}\sol\left(\chi_1; \p_1, \P_1\right),\\
Q_2^{\rm fin}&=\lambda_0\e^{\i\phi}\U_1
\sol\left(\chi_2;\p_2,\widetilde{\P}_2\right)\U_1^T.
\end{align}
Here, with $\Xi$ in eq.(\ref{shift}), we have introduced a deformed polarization matrix, 
\begin{align}
\widetilde{\P}_2	\hspace*{-1mm}=\hspace*{-1mm}\left(\hspace*{-2mm}
\begin{array}{cc}
\Xi \tr [\P_1\P_2]\hspace*{-2mm} &\hspace*{-2mm} 
\Xi^{1/2} \tr \left[ (-\i\sigma^y)\P_1\P_2\right]\\
\Xi^{1/2} \tr \left[ \P_1(-\i\sigma^y)^T\P_2\right] \hspace*{-1.5mm}&\hspace*{-1.5mm} 
\tr \left[ (-\i\sigma^y)\P_1(-\i\sigma^y)^T\P_2\right]
\end{array}
\hspace*{-2mm}\right).
\end{align}
In the presence of the soliton living at the right hand, 
the soliton living at the left hand has deformed boundary conditions. 
The deformed polarization matrix comes from the effects of the collision as well as 
the presence of another soliton at the right hand. 
The same is true for the ferro--ferro collision as will be seen later. 

\begin{figure}[htbp]
\begin{center}
\unitlength=1mm
\begin{picture}(150,75)
\put(18,0){(a)}
\put(56,0){(b)}
\put(94,0){(c)}
\put(132,0){(d)}
\put(18,72){$x$}
\put(56,72){$x$}
\put(94,72){$x$}
\put(132,72){$x$}
\put(38,30){$t$}
\put(76,30){$t$}
\put(115,30){$t$}
\put(153,30){$t$}
\put(-3,71){$|\phi_1|^2$}
\put(36,71){$|\phi_0|^2$}
\put(78,71){$n$}
\put(115,71){$f_x$}
\includegraphics[scale=0.47]{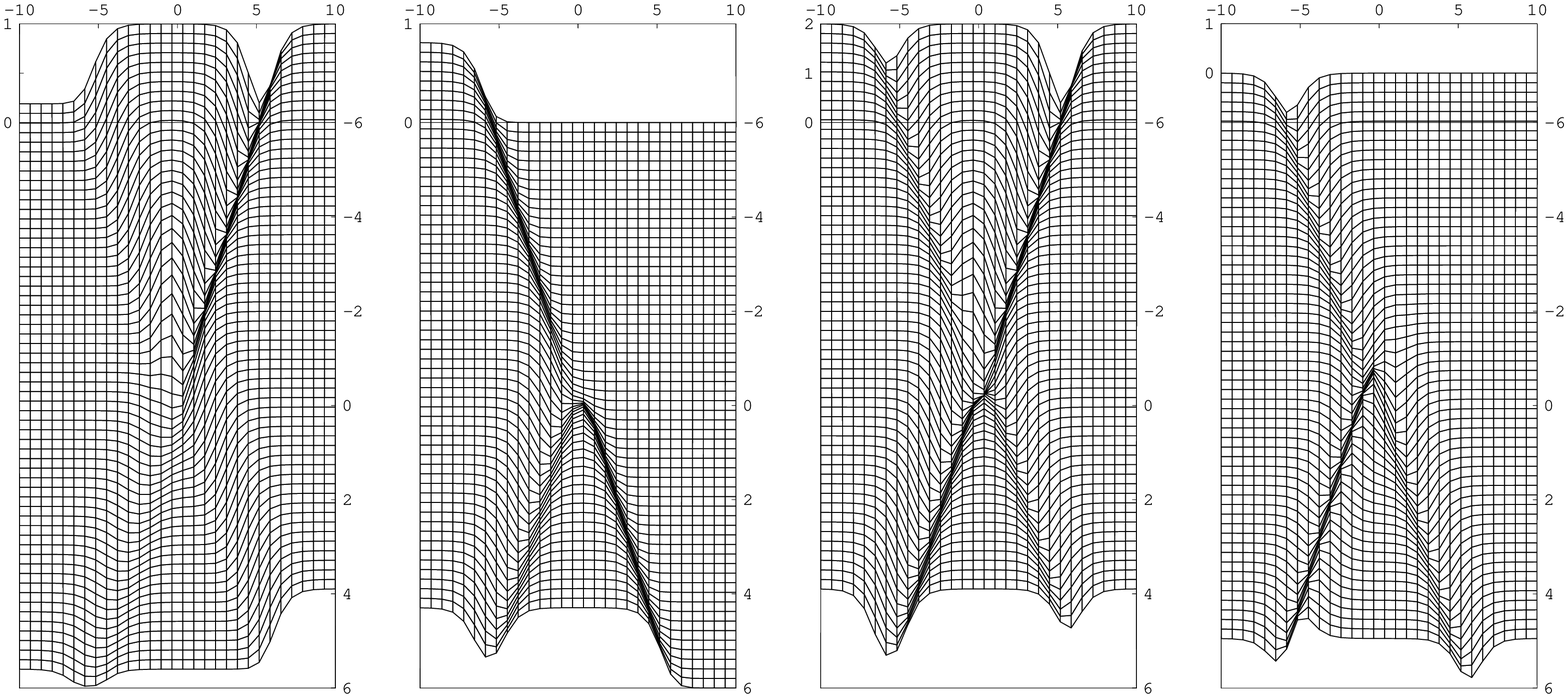}
\end{picture}
\end{center}
\caption{Density profiles of $2$-soliton collision. 
Soliton 1 trailing right-down is ferromagnetic and 
soliton 2 trailing left-down is polar. 
The parameters are fixed as $\lambda_0=1,\ k=0,\ \p_1=\pi-\p_2=1.12,\ 
\P_1=\left(\hspace*{-2mm}
\begin{array}{cc}
1\hspace*{-1mm}\vspace{-1mm}&\hspace*{-1mm}1\\ 
1\hspace*{-1mm}&\hspace*{-1mm}1
\end{array}
\hspace*{-2mm}\right),\ 
\P_2=\left(\hspace*{-2mm}
\begin{array}{cc}
1\hspace*{-1mm}\vspace{-1mm}&\hspace*{-1mm}0\\ 
0\hspace*{-1mm}&\hspace*{-1mm}1
\end{array}
\hspace*{-2mm}\right)$. 
(a) density $|\phi_{1}|^2$, (b) density $|\phi_0|^2$, 
(c) particle number density $n$, (d) spin density $f_x$. 
$|\phi_{-1}|^2$ behaves almost like $|\phi_1|^2$. 
$f_y$ and $f_z$ are zero everywhere. }
\label{fig:2solitonFP}
\end{figure}

Figure \ref{fig:2solitonFP} shows the behavior of the ferro--polar collision. 
Notice that the time direction is downward. 
It is observed for soliton 2 (polar) that much of the hole amplitude is moved from 
$\phi_{\pm1}$ to $\phi_0$ due to the collision. 
For soliton 1 (ferro), no such a spin-mixing (or spin-transfer) is observed and 
its domain-wall shape is preserved. 
This is explained as follows. 
Soliton 1 has nonzero spin so it can rotate the local spin density of soliton 2; 
soliton 2 has zero spin in total so it cannot affect the local spin density of soliton 1.
This kind of phenomenon was called a spin-switching 
in the attractive case \cite{IMW1, IMW2}. 
It is interesting to recognize that the spin density 
$f_x$ of soliton 2 shows up with a dipole-like shape after the collision 
(Fig.\ref{fig:2solitonFP}.(d)). 

\subsection{Ferromagnetic state vs. Ferromagnetic state}
Let us examine the collision of two solitons both in the ferromagnetic state. 
For $j=1, 2$, set $\det\P_j=0$, and write 
\begin{align}
\P_j=\exp\left[ -\i\theta_j \sigma^y/2\right] {\rm diag}(\Lambda_j,0)
\exp\left[ -\i\theta_j \sigma^y/2\right],
\label{Pi_specialization}
\end{align}
and 
\begin{align}
\U_j=\e^{\i\p_j} \exp\left[ -\i\theta_j \sigma^y/2\right] 
\exp\left[ \i\p_j\sigma^z/2\right].
\end{align}
Then we have the asymptotic forms as follows.  
From eq.(\ref{2soliton}), we have for $t\to-\infty$, 
\begin{align}
Q&\simeq Q_1^{\rm in}+Q_2^{\rm in}, \nonumber\\
Q_1^{\rm in}&=\lambda_0\e^{\i\phi}\U_2
\sol\left(\chi_1;\p_1,\widetilde{\P}_1\right)\U_2^T,\\
Q_2^{\rm in}&=\lambda_0\e^{\i\phi}\sol\left(\chi_2; \p_2, \P_2\right),
\end{align}
and for $t\to\infty$, 
\begin{align}
Q&\simeq Q_1^{\rm fin}+Q_2^{\rm fin}, \nonumber\\
Q_1^{\rm fin}&=\lambda_0\e^{\i\phi}\sol\left(\chi_1; \p_1, \P_1\right),\\
Q_2^{\rm fin}&=\lambda_0\e^{\i\phi}\U_1
\sol\left(\chi_2;\p_2,\widetilde{\P}_2\right)\U_1^T,
\end{align}
where deformed polarization matrices are obtained as 
\begin{align}
\widetilde{\P}_j
=\exp\left[ -\i\theta' \sigma^y/2\right] {\rm diag}(\Lambda,0)
\exp\left[ -\i\theta' \sigma^y/2\right],
\end{align}
for $j=1,2$. 
Here we have introduced new parameters, 
\begin{align}
&\Lambda=\Xi\cos^2\frac{1}{2}(\theta_2-\theta_1)+\sin^2\frac{1}{2}(\theta_2-\theta_1),\\
&\tan\frac{\theta'}{2}=\Xi^{-1/2}\tan\frac{1}{2}(\theta_2-\theta_1).
\end{align}
We check the boundary conditions for the asymptotic forms. 
We have for $t\to-\infty$, 
\begin{align}
Q_1^{\rm in}\e^{-\i\phi}\to \lambda_0\ \U_2\U_2^T
&,\qquad x\to\infty,\\
Q_1^{\rm in}\e^{-\i\phi}\to 
\lambda_0\ \U\U^T
&,\qquad x\to-\infty,\\
Q_2^{\rm in}\e^{-\i\phi}\to \lambda_0 I
&,\qquad x\to\infty,\\
Q_2^{\rm in}\e^{-\i\phi}\to 
\lambda_0\ \U_2\U_2^T
&,\qquad x\to-\infty,
\end{align}
and for $t\to\infty$, 
\begin{align}
Q_1^{\rm in}\e^{-\i\phi}\to \lambda_0 I
&,\qquad x\to\infty,\\
Q_1^{\rm in}\e^{-\i\phi}\to 
\lambda_0\ \U_1\U_1^T
&,\qquad x\to-\infty,\\
Q_2^{\rm in}\e^{-\i\phi}\to \lambda_0\ \U_1\U_1^T
&,\qquad x\to\infty,\\
Q_2^{\rm in}\e^{-\i\phi}\to 
\lambda_0\ \U\U^T
&,\qquad x\to-\infty,
\end{align}
where 
\begin{align}
\U=&\e^{\i(\p_1+\p_2)/2} \exp\left[-\i\theta_2\sigma^y/2\right] 
\exp\left[\i\p_2\sigma^z/2\right] \nonumber\\
&\times\exp\left[-\i\theta'\sigma^y/2\right] 
\exp\left[\i\p_1\sigma^z/2\right] \\
=&\e^{\i(\p_1+\p_2)/2} \exp\left[-\i\theta_1\sigma^y/2\right] 
\exp\left[\i\p_1\sigma^z/2\right] \nonumber\\
&\times\exp\left[-\i\theta'\sigma^y/2\right] 
\exp\left[\i\p_2\sigma^z/2\right].
\end{align}
%
%
%
\begin{figure}[tbp]
\begin{center}
\unitlength=1mm
\begin{picture}(150,75)
\put(18,-3){(a)}
\put(71,-3){(b)}
\put(127,-3){(c)}
\put(10,55){\vector(1,0){5}}
\put(26,55){\vector(-1,0){5}}
\put(8,20){\vector(-1,0){5}}
\put(32,20){\vector(1,0){5}}
\put(63,57){\vector(1,0){5}}
\put(80,57){\vector(-1,0){5}}
\put(60,24){\vector(-1,0){5}}
\put(83,24){\vector(1,0){5}}
\put(118,57){\vector(1,0){5}}
\put(136,57){\vector(-1,0){5}}
\put(113,22){\vector(-1,0){5}}
\put(139,22){\vector(1,0){5}}
\hspace*{-1.2cm}
\begin{minipage}[b]{53mm}
\includegraphics[scale=.4]{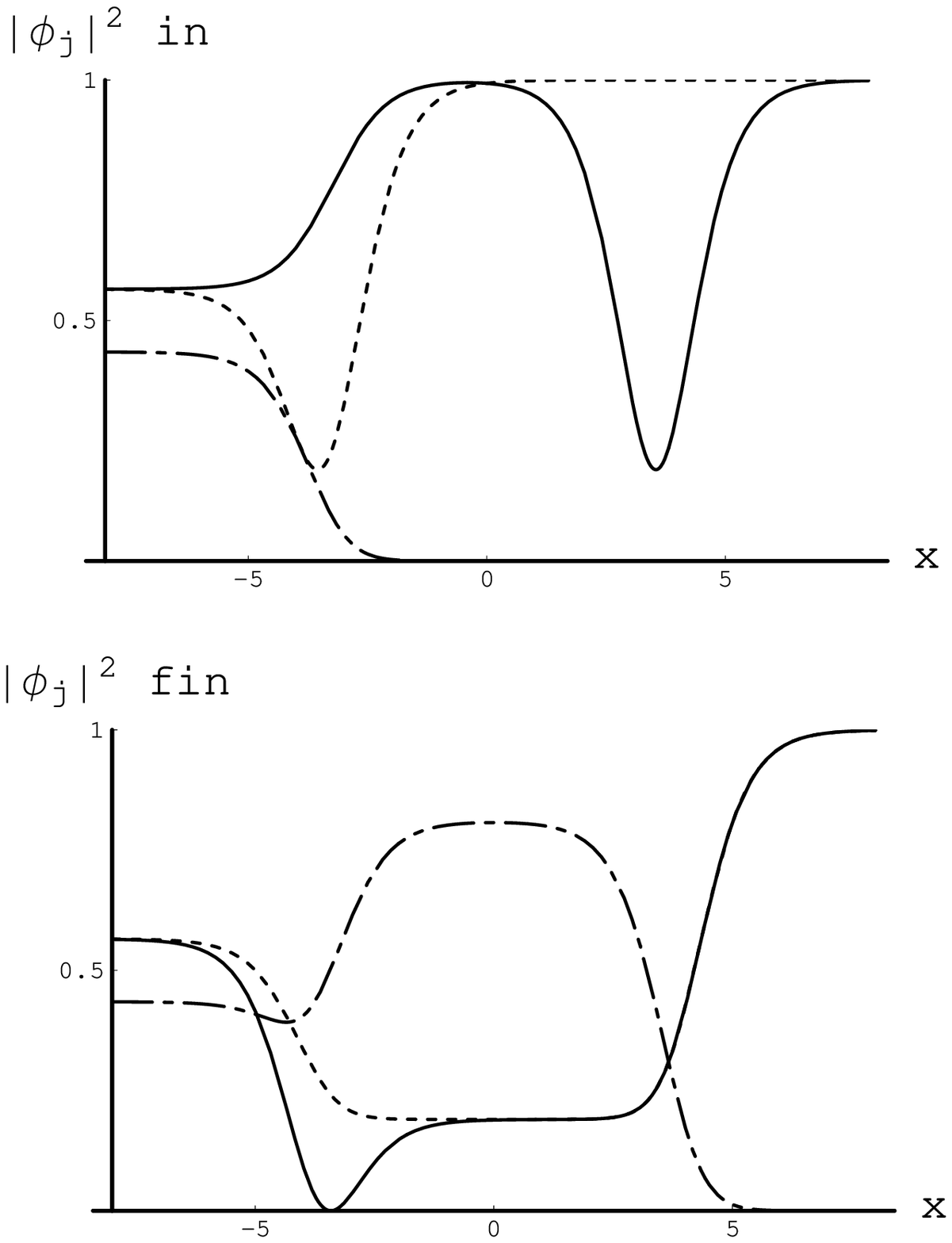}
\end{minipage}
\begin{minipage}[b]{53mm}
\includegraphics[scale=.4]{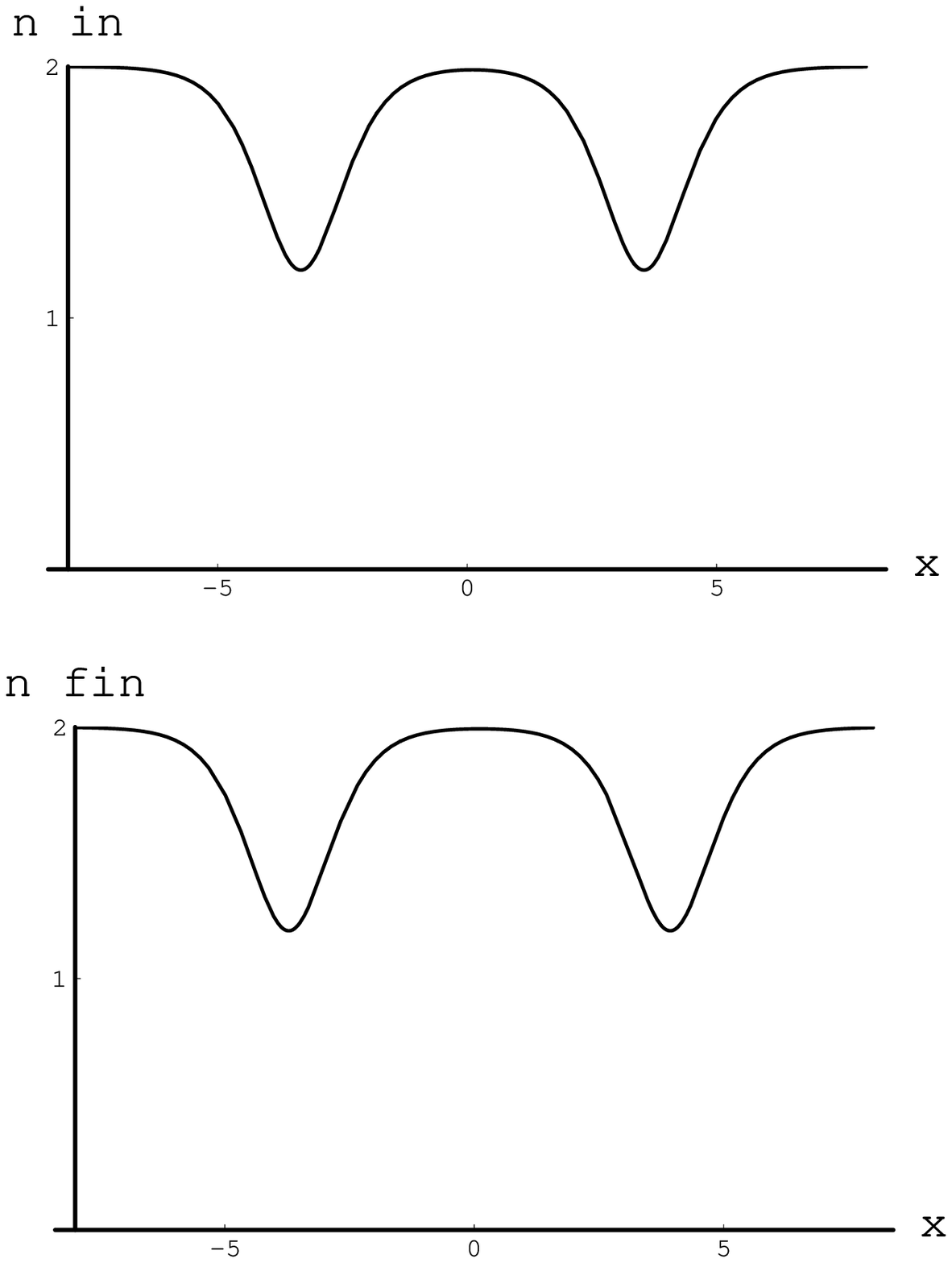}
\end{minipage}
\begin{minipage}[b]{53mm}
\includegraphics[scale=.4]{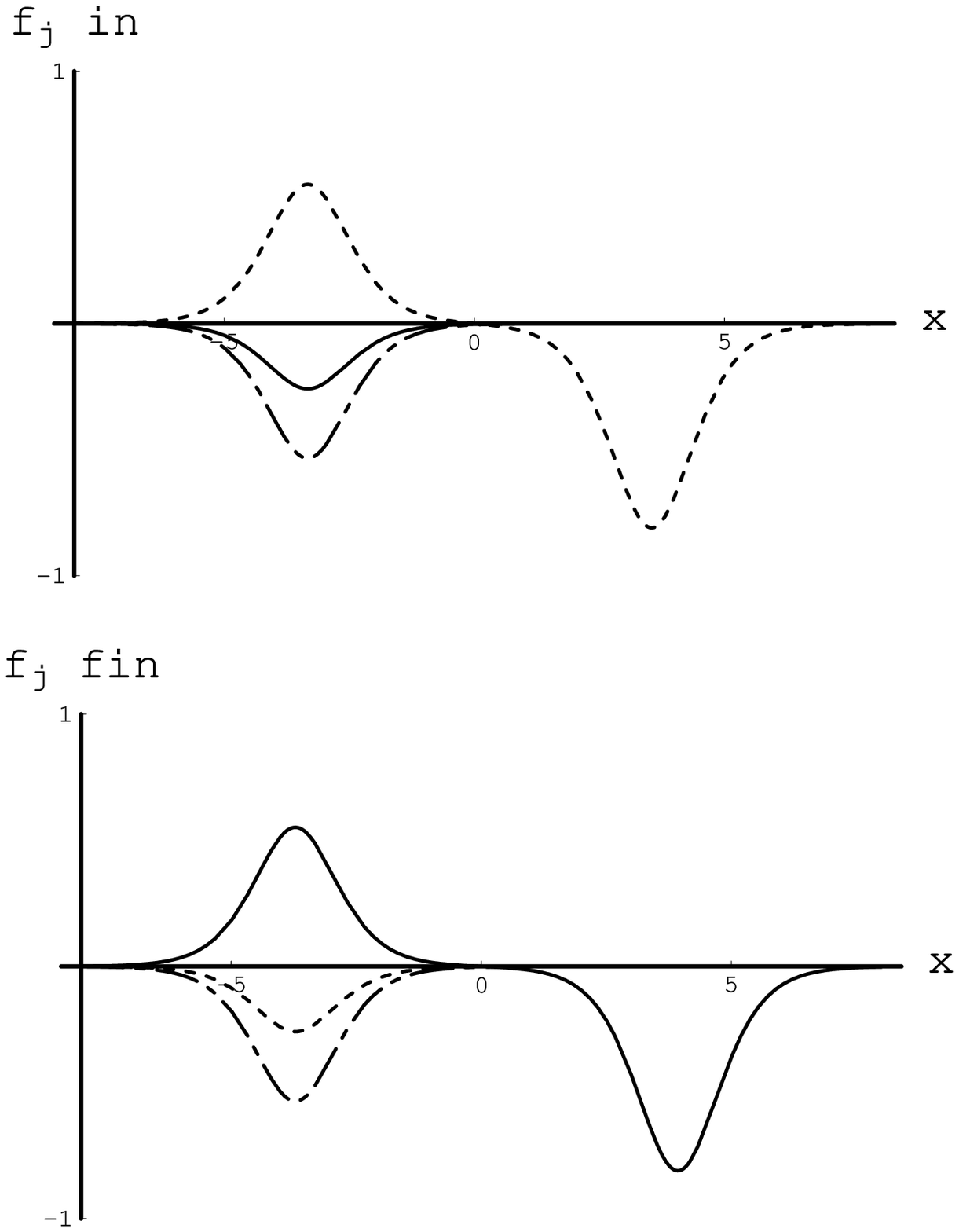}
\end{minipage}
\end{picture}
\caption{Snapshots before and after the collision of two solitons 
both in ferromagnetic state. 
The parameters are fixed as $\lambda_0=1,\ k=0,\ \p_1=\pi-\p_2=1.12,\ 
\P_1=\left(\hspace*{-2mm}
\begin{array}{cc}
1\hspace*{-1mm}\vspace{-1mm}&\hspace*{-1mm}1\\ 
1\hspace*{-1mm}&\hspace*{-1mm}1
\end{array}
\hspace*{-2mm}\right),\ 
\P_2=\left(\hspace*{-2mm}
\begin{array}{cc}
1\hspace*{-1mm}\vspace{-1mm}&\hspace*{-1mm}0\\ 
0\hspace*{-1mm}&\hspace*{-1mm}0
\end{array}
\hspace*{-2mm}\right)$. 
The collision takes place around $t=0$. 
The upper (lower) panels depict the behaviors 
before (after) the collision at $t=-4$ ($t=4$). 
(a) density of each component, $|\phi_1|^2$ (solid line), 
$|\phi_0|^2$ (chain line) and $|\phi_{-1}|^2$ (dotted line), (b) particle number density $n$, 
(c) spin densities, $f_x$ (solid line), $f_y$ (chain line) 
and $f_z$ (dotted line). 
Soliton 1 and soliton 2 have spins 
$\bm{F}_1^{\rm in}=(-0.575,-1.19,1.23)$, $\bm{F}_2^{\rm in}=(0,0,-1.8)$, 
$\bm{F}_1^{\rm fin}=(-1.8,0,0)$, $\bm{F}_2^{\rm fin}=(1.23, -1.19,-0.575)$. 
The rotation angle of spins around the total spin is $\omega=-1.38$. 
}\label{fig:FF}
\end{center}
\end{figure}

An example of the profiles of two solitons before and after the collision 
is shown in Fig. \ref{fig:FF}. 
The collision causes the spin precession. Let us see this phenomenon closely. 
Using eq.(\ref{spinrotation}), we find that 
the spins of the two solitons have the following form, 
\begin{align}
\label{spin-1}
\bm{F}_i^{\rm m}&=-|\bm{F}_i^{\rm m}| 
\frac{\tr\P_i^{\rm m} \bm{\sigma}}{\tr\P_i^{\rm m}},
\qquad |\bm{F}_i^{\rm m}|=2\lambda_0\sin\p_i ,
\end{align}
for $i=1,2$ and ${\rm m}={\rm in, fin}$. 
We find that the polarization matrices are factorized as 
\begin{align}
\label{spin-2}
\P_i^{\rm m}=\bm{u}_i^{\rm m}\cdot\bm{u}_i^{\rm m}{}^\dagger,
\end{align}
where 
\begin{align}
\label{u-1}
\bm{u}_i^{\rm m}=\e^{\i\p_j/2}\exp[-\i\theta_j\sigma^y/2]\exp[\i\p_j\sigma^z/2]
\left(\hspace*{-1mm}
\begin{array}{c}
\cos\frac{\theta'}{2}\\ \sin\frac{\theta'}{2}
\end{array}
\hspace*{-1mm}\right),
\end{align}
for $(i,j,{\rm m})=(1,2,{\rm in}),\ (2,1, {\rm fin})$, and 
\begin{align}
\label{u-2}
\bm{u}_i^{\rm m}&=\left(\hspace*{-1mm}
\begin{array}{c}
\cos\frac{\theta_i}{2}\\
\sin\frac{\theta_i}{2}
\end{array}
\hspace*{-1mm}\right),
\end{align}
for $(i,{\rm m})=(1,{\rm fin}),\ (2,{\rm in})$. 
Explicitly, the spins are calculated as 
\begin{align}
\bm{F}_i^{\rm m}=-|\bm{F}_i^{\rm m}|\left(\hspace*{-1mm}
\begin{array}{c}
\cos\theta_j\cos\p_j\sin\theta'+\sin\theta_j\cos\theta'\\
-\sin\p_j\sin\theta'\\
-\sin\theta_j\cos\p_j\sin\theta'+\cos\theta_j\cos\theta'
\end{array}
\hspace*{-1mm}\right),
\end{align}
for $(i,j,{\rm m})=(1,2,{\rm in}),\ (2,1, {\rm fin})$, and 
\begin{align}
\bm{F}_i^{\rm m}=-|\bm{F}_i^{\rm m}|\left(\hspace*{-1mm}
\begin{array}{c}
\sin\theta_i\\
0\\
\cos\theta_i
\end{array}
\hspace*{-1mm}\right),
\end{align}
for $(i,{\rm m})=(1,{\rm fin}),\ (2,{\rm in})$. 
Also, for the total spin, we have 
\begin{align}
|\bm{F}_T|=2\lambda_0(\sin^2\!\p_1+\sin^2\!\p_2+2\sin\p_1\sin\p_2\cos\theta')^{1/2}.
\end{align}
From eqs.(\ref{u-1}) and (\ref{u-2}), one can show the relation, 
\begin{align}
\label{u-3}
\bm{u}_i^{\rm m}=-\Lambda\left(\bm{u}_i^{\rm n}-(1+\e^{-\i\p_j}\Xi^{-1/2})
(\bm{u}_j^{\rm n}{}^\dagger\cdot \bm{u}_i^{\rm n})\bm{u}_j^{\rm n}\right),
\end{align}
for $(i,j,{\rm m},{\rm n})=(1,2,{\rm fin},{\rm in}),\ (2,1,{\rm in},{\rm fin})$. 
Using the identity, 
\begin{align}
\tr[ (\bm{u}_2^\dagger\cdot \bm{u}_1)\bm{u}_2\cdot\bm{u}_1^\dagger
\bm{\sigma}]
=\frac{1}{2}\left[\bm{F}_1+\bm{F}_2+\i(\bm{F}_2\times\bm{F}_1)\right],
\end{align}
for $\bm{F}_i=\tr[\bm{u}_i\cdot\bm{u}_i^\dagger\bm{\sigma}]$ 
together with eqs.(\ref{spin-1}) and (\ref{spin-2}), we arrive at 
a formula for the rotation of spins, 
\begin{align}
\bm{F}_j^{\rm fin}=&\frac{1}{2}\Big\{
[(1+\Delta_{jl})+(1-\Delta_{jl})\cos\omega] \bm{F}_j^{\rm in}\nonumber\\
&\quad+[(1+\Delta_{jl})-(1+\Delta_{jl})\cos\omega] \bm{F}_l^{\rm in}\nonumber\\
&\quad+2\sin\omega\frac{\bm{F}_j^{\rm in}\times\bm{F}_l^{\rm in}}{|\bm{F}_T|}\Big\},
\end{align}
for $(j,l)=(1,2),(2,1)$ where 
\begin{align}
\Delta_{jl}=\frac{|\bm{F}_j|^2-|\bm{F}_l|^2}{|\bm{F}_T|^2}.
\end{align}
The rotation angle $\omega$ around the total spin $\bm{F}_T$ is determined by 
\begin{align}
\sin\omega=-\frac{\lambda_0\Xi^{1/2}|\bm{F}_T|}
{(1+\Xi)+(1-\Xi)\bm{F}_1\cdot\bm{F}_2/|\bm{F}_1| |\bm{F}_2|}.
\end{align}
Recall that $\Xi$ is defined in eq.(\ref{shift}).

\section{Conclusion}

In this paper, we have studied soliton solutions of the integrable multi-component 
Gross--Pitaevskii equation for $F=1$ spinor Bose--Einstein condensate (BEC) 
with repulsive and anti-ferromagnetic 
interactions. The solutions are directly obtained by identification with those of 
the self-defocusing matrix nonlinear Schr\"{o}dinger equation (NLSE), 
which was solved by the inverse scattering method (ISM) in ref. \citen{IUW}. 
Since the interactions are repulsive, the solitons arise as dark solitons in general 
under nonvanishing boundary conditions. 
These dark solitons have similar properties 
compared to bright solitons in $F=1$ spinor BEC 
with attractive and ferromagnetic interactions \cite{IMW1, IMW2}. 

We have shown that 
one-solitons are either in the ferromagnetic state or in the polar state. 
A soliton in the polar state is proved to be a special case of 
two solitons in the ferromagnetic state moving with the same velocity. 
These two states, ferromagnetic or polar, 
are selected by choosing the boundary conditions. 
When one takes the left and right boundary amplitudes
of wavefunctions differently, i.e., the SU(2) ``rotated" boundary conditions, 
the one-soliton is in the ferromagnetic state. 
On the other hand, when the boundary conditions differ only by phase, 
we get the polar state. 
As is always the case for dark solitons, these changes in the left and right boundary conditions make solitons the topological objects (excitations), guaranteeing the stabilities and the existence of finite spin amplitudes even in the anti-ferromagnetic interactions. 
Two-soliton collisions for every combination of the states have been examined. 
A soliton in the ferromagnetic state, which has nonzero total spin, 
causes spin rotation of another soliton by the collision. 
However, a soliton in the polar state do not rotate the spin of another soliton 
because it has zero spin in total. 
It can be said that, as ``magnetic" carriers, 
solitons in the ferromagnetic state are operative while 
those in the polar state are passive. 

To detail the $N$-soliton collisions is straightforward. 
As is confirmed in the asymptotic forms of the two-soliton solution, 
when the other solitons are far apart, a soliton keeps the form of one-soliton 
with deformed boundary conditions and a deformed polarization matrix, affected by 
the solitons living at the right hand side of it. 
The explicit expression of the formula for multi-soliton collisions 
may be complicated and needs a further analysis. 
As investigated for the Manakov model in ref. \citen{T}, 
it is important to clarify the factorization property 
of the $N$-soliton collision. 

Very recently, plane wave solutions and soliton solutions are obtained 
for general coupling constants $\bar{c}_0$ and $\bar{c}_2$ \cite{WT06}. 
The stability of the solitons is numerically studied \cite{Li}. 
An interesting question is whether the multi-component Gross--Pitaevskii equation is 
integrable only at some special points as considered here. 
In particular, its generalization into the generic hyperfine spin $F$ case is 
very interesting \cite{IMW3} and is left for a future problem.


\vspace{2cm}
\appendix
\section{Explicit Expression for Two-Soliton Solution}

We give the explicit form of $2$-soliton solution of the matrix NLSE (\ref{NLS}) 
as follows. 
\begin{align}
\label{2soliton}
Q(x,t)=\lambda_0 \e^{\i\phi(x,t)}
\frac{\mathfrak{B}}{\mathfrak{A}},
\end{align}
where with $\Xi$ in eq.(\ref{shift}) and the polarization matrices $\P_j$, 
\begin{align*}
\mathfrak{A}=&
\Xi^2 \det\P _1\det\P _2 \e^{2(\chi_1+\r_1)+2(\chi_2+\r_2)}
+\Xi \det\P _1\tr\P _2 \e^{2(\chi_1+\r_1)+(\chi_2+\r_2)}\\
&+\det\P _1\e^{2(\chi_1+\r_1)}
+\Xi \tr\P _1\det\P _2 \e^{(\chi_1+\r_1)+2(\chi_2+\r_2)}\\
&+\left[ \e^{\r_1+\r_2}\tr\P _1\tr\P _2-\sin^{-2}\left(\frac{1}{2}(\p_1+\p_2)\right)
\tr\P _1\P _2\right] \e^{\chi_1+\chi_2}\\
&+\tr\P _1 \e^{\chi_1+\r_1}
+\det\P _2 \e^{2(\chi_2+\r_2)}+\tr\P _2\e^{\chi_2+\r_2}+1,
\end{align*}
and $\mathfrak{B}=(\mathfrak{B}_{ij})$ is a $2\times 2$ matrix such that 
\begin{align*}
\mathfrak{B}_{11}=&
\Xi^2 \e^{2\i(\p_1+\p_2)}\det\P_1 \det\P_2 \e^{2(\chi_1+\r_1)+2(\chi_2+\r_2)}\\
&+\Xi \det\P_1 \left[ \e^{2\i(\p_1+\p_2)} (\P_2)_{11}+\e^{2\i\p_1} (\P_2)_{22}\right]
\e^{2(\chi_1+\r_1)+(\chi_2+\r_2)}+\e^{2\i\p_1} \det\P_1 \e^{2(\chi_1+\r_1)}\\
&+\Xi \det\P _2\left[ \e^{2\i(\p_1+\p_2)} (\P_1)_{11}+\e^{2\i\p_2} (\P_1)_{22}\right]
 \e^{(\chi_1+\r_1)+2(\chi_2+\r_2)}\\
&+\Big\{ \Xi \e^{2\i(\p_1+\p_2)}(\P_1)_{11}(\P_2)_{11}+\Xi (\P_1)_{22}(\P_2)_{22}
+\e^{2\i\p_1} (\P_1)_{11}(\P_2)_{22}+\e^{2\i\p_2} (\P_1)_{22}(\P_2)_{11}\\
&\qquad
-\sin^{-2}\left(\frac{1}{2}(\p_1+\p_2)\right) \e^{-(\r_1+\r_2)+\i(\p_1+\p_2)}
\left[(\P_1)_{12} (\P_2)_{21}+(\P_1)_{21} (\P_2)_{12}\right] \Big\}
\e^{(\chi_1+\r_1)+(\chi_2+\r_2)}\\
&+\left[ \e^{2\i\p_1} (\P_1)_{11} +(\P_1)_{22}\right] \e^{\chi_1+\r_1}\\
&+\e^{2\i\p_2} \det\P_2 \e^{2(\chi_2+\r_2)}
+\left[ \e^{2\i\p_2} (\P_2)_{11} +(\P_2)_{22}\right] \e^{\chi_2+\r_2}+1,
\\
\vspace{1cm}\\
\mathfrak{B}_{12}=&
2\i\Big\{
\Xi \e^{2\r_1+2\i\p_1+\i\p_2} \det\P_1 (\P_2)_{12} \e^{2\chi_1+\chi_2}
+\Xi \e^{2\r_2+\i\p_1+2\i\p_2} (\P_1)_{12} \det\P_2 \e^{\chi_1+2\chi_2}\\
&+\left[-\Xi^{1/2}\e^{\r_2+\i\p_1+\i\p_2} (\P_1)_{12} \tr\P_2 
+\Xi^{1/2}\e^{\r_1+\i\p_1+\i\p_2}\tr\P_1 (\P_2)_{12}\right] \e^{\chi_1+\chi_2}\\
&+\e^{\i\p_1} (\P_1)_{12} \e^{\chi_1}+\e^{\i\p_2} (\P_2)_{12} \e^{\chi_2}\Big\} . 
\end{align*}
We note that $\mathfrak{B}_{21}$ (resp. $\mathfrak{B}_{22}$) is given by 
replacing the indices of $\P_j$ with $1\leftrightarrow 2$ 
in $\mathfrak{B}_{12}$ (resp. $\mathfrak{B}_{11}$). 

When we choose $\p_1=\p_2=\p$ for two ferromagnetic solitons where 
$\det\P_1=\det\P_2=0$, we have a one-soliton solution (\ref{eq:1soliton}), 
\begin{align*}
Q(x,t)=\lambda_0\e^{\i\phi(x,t)}\sol\left(\chi;\p,\P\right), 
\end{align*}
where $\chi_1=\chi_2=\chi$ and $\P=\P_1+\P_2$. 
This one-soliton is in the polar state if $\det\P\neq0$.  
Remark that $\det\P_1=0$ and $\det\P_2=0$ do not make $\det\P=0$ for 
$\P=\P_1+\P_2$ in general. 
This kind of reduction gives multi-solitons in ferromagnetic and polar states. 
Then, we can consider (\ref{2soliton}) with $\det\P_j\ge0$ and $\tr\P_j>0$ 
($j=1, 2$) as the general form of $2$-soliton solution. 


\begin{thebibliography}{99} 
\bibitem{Stenger}
J. Stenger, S. Inouye, D. M. Stamper-Kurn, H.-J. Miesner, A. P. Chikkatur and 
W. Ketterle: 
Nature (London) \textbf{396} (1998) 345. 
\bibitem{Stamper}
C. M. Stamper-Kurn, M. R. Andrews, A. P. Chikkatur, S. Inouye, H.-J. Miesner, 
J. Stenger and W. Ketterle: Phys. Rev. Lett. \textbf{80} (1998) 2027. 
\bibitem{Miesner}
H.-J. Miesner, D. M. Stamper-Kurn, J. Stenger, S. Inouye, A. P. Chikkatur and 
W. Ketterle: Phys. Rev. Lett. \textbf{82} (1999) 2228.
%
\bibitem{Meystre}
P. Meystre, \textit{Atom Optics} (Springer-Verlag, New York, 2001). 
\bibitem{Burger}
S. Burger, K. Bongs, S. Dettmer, W. Ertmer and K. Sengstock: 
Phys. Rev. Lett. \textbf{83} (1999) 5198.
\bibitem{Denschlag}
J. Denschlag, J. E. Simsarian, D. L. Feder, C. W. Clark, L. A. Collins, 
J. Cubizolles, L. Deng, E. W. Hagley, K. Helmerson, W. P. Reinhardt, 
S. L. Rolston, B. I. Schneider and W. D. Phillips: 
Science \textbf{287} (2000) 97. 
\bibitem{Strecker}
K. E. Strecker, G. B. Partridge, A. G. Truscott and R. G. Hulet: Nature (London) 
\textbf{417} (2002) 150.
\bibitem{Khaykovich}
L. Khaykovich, F. Schreck, G. Ferrari, T. Bourdel, J. Cubizolles, L. D. Carr, 
Y. Castin and C. Salomon: Science \textbf{296} (2002) 1290.
\bibitem{IMW1}
J. Ieda, T. Miyakawa and M. Wadati: Phys. Rev. Lett. \textbf{93} (2004) 194102. 
\bibitem{IMW2}
J. Ieda, T. Miyakawa and M. Wadati: J. Phys. Soc. Jpn. \textbf{73} (2004) 2996. 
\bibitem{TW1}
T. Tsuchida and M. Wadati: J. Phys. Soc. Jpn. \textbf{67} (1998) 1175. 
\bibitem{T}
T. Tsuchida: Prog. Theor. Phys. \textbf{111} (2004) 151. 
\bibitem{IUW}
J. Ieda, M. Uchiyama and M. Wadati: arXiv:nlin.SI/0603010. 
\bibitem{WT06}
M. Wadati and N. Tsuchida: J. Phys. Soc. Jpn. \textbf{75} (2006) 014301. 
\bibitem{Li}
L. Li, Z. Li, B. A. Malomed, D. Mihalache and W. M. Liu: Phys. Rev. A \textbf{72} 
(2005) 033611.
\bibitem{IMW3}
J. Ieda, T. Miyakawa and M. Wadati: to be published in Laser Physics \textbf{16} (2006).
\end{thebibliography}
\end{document}